\documentclass[a4paper,english,10pt]{article}
\usepackage[T1]{fontenc}
\usepackage[latin9]{inputenc}
\usepackage{color}
\usepackage{amsmath}
\usepackage{amssymb}
\usepackage{esint}
\usepackage{harvard}

\usepackage{babel} 

\begin{document}

\title{Thermal properties of materials from ab-initio quasi-harmonic phonons}
\author{
  Stefano Baroni \\
  SISSA -- Scuola Internazionale Superiore di Studi Avanzati, and \\
  CNR DEMOCRITOS National Simulation Center, Trieste   \\ \\
  Paolo Giannozzi \\
  Dipartimento di Fisica, Universit\`a di Udine, Udine, and \\
  CNR DEMOCRITOS National Simulation Center, Trieste \\ \\ and \\ \\
  Eyvaz Isaev \\
  Department of Physics, Chemistry and Biology (IFM), \\
  Link\"oping University, Sweden, and \\
  Theoretical Physics Department, \\
  Moscow State Institute of Steel and
  Alloys, Russia 
}

\maketitle

\section{Introduction}

Computer simulations allow for the investigation of many materials
properties and processes that are not easily accessible in the
laboratory.  This is particularly true in the Earth sciences, where
the relevant pressures and temperatures may be so extreme that no
experimental techniques can operate at those
conditions. Computer modeling is often the only
source of information on the properties of materials that, combined
with indirect evidence (such as \emph{e.g.} seismic data), allows one
to discriminate among competing planetary models. Many computer
simulations are performed using effective inter-atomic potentials
taylored to reproduce some experimentally observed properties of the
materials being investigated.  The remoteness of the physically
interesting conditions from those achievable in the laboratory, as
well as the huge variety of different atomic coordination and local
chemical state occurring in the Earth interior, make the dependability
of semi-empirical potentials questionable.
First-principles techniques based on density-functional theory
(DFT) \cite{HK:64,KS:65} are much more predictive, not being biased by
any prior experimental input, and have demonstrated a considerable
accuracy in a wide class of materials and variety of external
conditions.  The importance of thermal effects in the range of
phenomena interesting to the Earth sciences makes a proper account of
atomic motion essential.  Traditionally, this is achieved using
molecular dynamics techniques which have been successfully combined
with DFT in the first-principles molecular dynamics technique of Car
and Parrinello \cite{CarParr}. Well below the melting temperature, the
numerical efficiency of molecular dynamics is limited by the lack of
ergodicity, which would require long simulation times, and by the
importance of long-wavelength collective motions (phonons), which
would require large simulation cells. Both difficulties are
successfully dealt with in the quasi-harmonic approximation (QHA) where
the thermal properties of solid materials are traced back to those of
a system of non-interacting phonons (whose frequencies are however
allowed to depend on volume or on other thermodynamic constraints). An
additional advantage of the QHA is that it accounts for
quantum-mechanical zero-point effects, which would not be accessible
to molecular dynamics with classical nuclear motion. The availability of suitable
techniques to calculate the vibrational properties of extended
materials using a combination of DFT and linear-response techniques
(resulting in the so-called density-functional perturbation theory,
DFPT \cite{BGT:87,Baroni-RMP:01}) makes it possible to combine the QHA
with DFT. The resulting simulation methodology has shown to be
remarkably accurate in a wide temperature range, extending up to not
very far from the melting line and has been applied to a wide variety
of systems, including many which are relevant to the Earth
sciences. This paper gives a short overview of the calculation of
thermal properties of materials in the framework of the QHA, using
DFT. The paper is organized as follows: in Sec. 2 we introduce some of
the thermal properties of interest and describe how they can be
calculated in the framework of the QHA; in Sec. 3 we describe the DFPT
approach to lattice dynamics; in Sec. 4 we briefly introduce some of
the computer codes that can be used to perform this task; in Sec. 5 we
review some of the application of the first-principles QHA to the
study of the thermal properties of materials; finally, Sec. 6 contains
our conclusions.

\section{Thermal properties and the Quasi-Harmonic
 Approximation} \label{sec:QHA}

The low-temperature specific heat of solids is 
experimentally found to vanish as the cube of the temperature,
with a cubic coefficient that is system-specific \cite{kittel,wallace}.
This is contrary to the predictions of classical statistical mechanics,
according to which the heat capacity of a system of harmonic oscillators
does not depend on temperature, nor on its spectrum. One of the landmarks
of modern solid-state physics, that greatly contributed to the establishment
of our present quantum-mechanical picture of matter, is the Debye
model for the heat capacity of solids. This model naturally explains
the low-temperature specific heat of solids in terms of the (quantum)
statistical mechanics of an ensemble of harmonic oscillators, which
can in turn be pictorially described as a gas of non-interacting quasi-particles
obeying the Bose-Einstein statistics (\emph{phonons}).

The internal energy of a single harmonic oscillator of angular frequency
$\omega$, in thermal equilibrium at temperature $T$, is:
\begin{equation}
 \langle E\rangle=\frac{\hbar\omega}{2}+
 \frac{\hbar\omega}{\mathrm{e^{\frac{\hbar\omega}{k_{B}T}}}-1},
 \label{eq:1oscillator-energy}
\end{equation}
where $k_{B}$ is the Boltzmann constant. By differentiating with
respect to temperature the sum over all the possible values of the
phonon momentum in the Brillouin zone (BZ) of
Eq. \eqref{eq:1oscillator-energy}, the constant-volume specific heat
of a crystal reads:
\begin{eqnarray} C_{V}(T) & = &
 \frac{1}{V}\sum_{\mathbf{q}\nu}\hbar\omega(\mathbf{q},\nu)n'(\mathbf{q},\nu),
 \label{eq:harmonic-specific-heat}
\end{eqnarray}
where $\omega(\mathbf{q},\nu)$ is the frequency of the $\nu$-th mode
(phonon) at point $\mathbf{q}$ in the BZ,
$n'(\mathbf{q,}\nu)=\frac{\partial}{\partial T}
\left(\mathrm{e}^{\frac{\hbar\omega(\mathbf{q},\nu)}{k_{B}T}}-1\right)^{-1}$,
and the sum is extended to the first BZ. By assuming that there are
three degenerate modes at each point of the BZ, each one with
frequency $\omega(\mathbf{q},\nu)=c|\mathbf{q}|$, $c$ being the sound
velocity, and converting the sum in
Eq. \eqref{eq:harmonic-specific-heat} into an integral, the resulting
expression for the heat capacity, valid in the low-temperature limit,
reads:
\begin{equation}
 C_{V}(T)=\frac{1}{\Omega}\frac{12\pi^{4}}{5}k_{B}
 \left(\frac{T}{\Theta_{D}}\right)^{3},\label{eq:cv-debye}
\end{equation}
where $\Omega$ is the volume of the crystal unit cell and
$\Theta_{D}=\frac{2\pi\hbar}{k_{B}}c\left(\frac{3}{4\pi\Omega}\right)^{\frac{1}{3}}$
is the so-called Debye temperature.

In the Born-Oppenheimer approximation \cite{Born-Oppenheimer}, the
vibrational properties of molecules and solids are determined by their
electronic structure through the dependence of the ground-state energy
on the coordinates of the atomic nuclei \cite{martin-book}. At low
temperature the amplitudes of atomic vibrations are much smaller than
inter-atomic distances, and one can assume that the dependence of
the ground-state energy on the deviation from equilibrium of the atomic
positions is quadratic. In this, so called \emph{harmonic}, approximation
(HA) energy differences can be calculated from electronic-structure
theory using static response functions \cite{P.D.DeCicco04291969,PhysRevB.1.910}
or perturbation theory \cite{BGT:87,Baroni-RMP:01} (See Sec. \ref{sec:ab-initio}).

In the HA, vibrational frequencies do not depend on interatomic
distances, so that the vibrational contribution to the crystal
internal energy does not depend on volume. As a consequence,
constant-pressure and constant-volume specific heats coincide in this
approximation, and the equilibrium volume of a crystal does not depend
on temperature.  Other shortcomings of the HA include its prediction
of an infinite thermal conductivity, infinite phonon lifetimes, and
the independence of vibrational spectra (as well as related
properties: elastic constants, sound velocities etc.) on temperature,
to name but a few. A proper account of anharmonic effects on the
static and dynamical properties of materials would require the
calculation of phonon-phonon interaction coefficients for all modes in
the BZ. Although the leading terms of such interactions can be
computed even from first principles
\cite{SolStComm.91.813,PhysRevLett.75.1819}---and the resulting
vibrational linewidths have in fact been evaluated in some cases
\cite{PhysRevLett.75.1819,PhysRevB.68.220509,PhysRevLett_99_176802}---the
extensive sampling of the phonon-phonon interactions over the BZ
required for free-energy evaluations remains a daunting task. The
simplest generalization of the HA, which corrects for most of the
above mentioned deficiencies, while not requiring any explicit
calculation of anharmonic interaction coefficients, is the QHA.

In the QHA, the crystal free energy is assumed to be determined by
the vibrational spectrum via the standard harmonic expression:
\begin{equation}
F(X,T)=U_{0}(X)+\frac{1}{2}\sum_{\mathbf{q}\nu}\hbar
\omega(\mathbf{q},\nu|X)+k_{B}T \sum_{\mathbf{q}\nu} 
\log\left(1-\mathrm{e}^{-\frac{\hbar\omega(\mathbf{q},\nu|X)}{k_{B}T}}\right),
\label{eq:qha-F}
\end{equation}
where $X$ indicates any global static constraint upon which
vibrational frequencies may depend (most commonly just the volume $V$,
but $X$ may also include anisotropic components of the strain tensor,
some externally applied fields, the internal distortions of the
crystal unit cell, or other thermodynamic constraints that may be
applied to the system), and $U_{0}(X)$ is the zero-temperature energy
of the crystal as a function of $X$. In the case $X=V$,
differentiation of Eq. \eqref{eq:qha-F} with respect to volume gives
the equation of state:
\begin{eqnarray}
 P & = & -\frac{\partial F}{\partial V}\nonumber \\
 & = & -\frac{\partial U_{0}}{\partial V}+
\frac{1}{V}\sum_{\mathbf{q}\nu}\hbar\omega(\mathbf{q},\nu)
\gamma(\mathbf{q},\nu)\left(\frac{1}{2}+
\frac{1}{\mathrm{e}^{\frac{\hbar\omega(\mathbf{q},\nu)}{k_{B}T}}-1}\right),
\label{eq:qha-eq_of_state}\end{eqnarray}
where 
\begin{equation}
 \gamma(\mathbf{q},\nu)=-\frac{V}{\omega(\mathbf{q},\nu)}
 \frac{\partial\omega(\mathbf{q},\nu)}{\partial
   V}\label{eq:grueneisen}
\end{equation} 
are the so-called Gr\"uneisen mode parameters. In a perfectly harmonic
crystal, phonon frequencies do not depend on the interatomic
distances, hence on volume. In such a harmonic crystal
Eq. \eqref{eq:qha-eq_of_state} implies that the temperature derivative
of pressure at fixed volume vanish: $(\partial P/\partial
T)_{V}=0$. It follows that the thermal expansivity,
$\beta=V^{-1}(\partial V/\partial T)_{P}$, which is given by the
thermodynamical relation:\begin{eqnarray}
 \beta & = & -\frac{(\partial P/\partial T)_{V}}
 {(\partial P/\partial V)_{T}}\label{eq:expansivity-1}\\
 & = & \frac{1}{B_{T}}\left(\frac{\partial P}{\partial T}\right)_{V}\label{eq:expansivity-2}\\
 & = &
 \frac{1}{B_{T}}\sum_{\mathbf{q}\nu}\hbar\omega(\mathbf{q},\nu)\gamma(\mathbf{q},
 \nu)n'(\mathbf{q},\nu),\label{eq:expansivity-3}\end{eqnarray} 
 where \textbf{\emph{ $B_{T}$=V($\partial P/\partial V)_{T}$}} 
is the crystal bulk modulus, would also vanish for perfectly harmonic
crystals.  Inspired by Eq. \eqref{eq:harmonic-specific-heat}, let us
define
$C_{V}(\mathbf{q,}\nu)=\hbar\omega(\mathbf{q},\nu)n'(\mathbf{q},\nu)/V$
as the contribution of the $\nu$-th normal mode at the \textbf{q}
point of the BZ to the total specific heat, and $\gamma$ as the
weighted average of the various Gr\"uneisen parameters:\begin{equation}
 \gamma=\frac{\sum_{\mathbf{q}\nu}\gamma(\mathbf{q},\nu)C_{V}(\mathbf{q},\nu)}{\sum_{\mathbf{q}\nu}C_{V}(\mathbf{q},\nu)}.\label{eq:total-Grueneisen}\end{equation}
In terms of $\gamma$, the thermal expansivity simply
reads:\begin{equation} \beta=\frac{\gamma
   C_{V}}{B_{T}}.\label{eq:compact-expansivity}\end{equation} The
vanishing of the thermal expansivity in the HA would also imply the
equality of the constant-pressure and constant-volume specific
heats. By imposing that the total differentials of the entropy as a
function of pressure and temperature or of volume and temperature
coincide, and by using the Maxwell identities, one can in fact show
that \cite{wallace}:

\begin{eqnarray}
C_{P}-C_{V} & = & -\frac{T}{V}\left(\frac{\partial P}{\partial V}\right)_{T}\left(\frac{\partial V}{\partial T}\right)_{P}\\
& = & TB_{T}\beta^{2}.\end{eqnarray}

We conclude this brief introduction to the QHA by noticing that the
ansatz given by Eq. \eqref{eq:qha-F} for the crystal free energy
in terms of its (volume-dependent) vibrational frequencies gives immediate
access to all the equilibrium thermal properties of the system. Whether
this \emph{implicit} account of anharmonic effects through the volume
dependence of the vibrational frequency only is sufficient to describe
the relevant thermal effects, or else an \emph{explicit} account of
the various phonon-phonon interactions is in order, instead, is a
question that can only be settled by extensive numeric experience.

\section{Ab-initio phonons} \label{sec:ab-initio}

\subsection{Lattice dynamics from electronic-structure
 theory} \label{sec:ld-from-est}

Several simplified approaches exist that allow to calculate full
(harmonic) phonon dispersions $\omega(\mathbf{q},\nu)$ from
semi-empirical force fields or inter-atomic potentials
\cite{Bruesch,Singh_Phys_Rep}.  The accuracy of such semi-empirical
models is however often limited to the physical conditions (pressure,
atomic coordination, crystal structure, etc.) at which the
inter-atomic potentials are fitted.  Really predictive calculations,
not biased by the  experimental information used to describe
inter-atomic interactions require a proper quantum-mechanical
description of the chemical bonds that held matter together. This can
be achieved in the framework of electronic-structure theory
\cite{martin-book}, starting from the \emph{adiabatic }or Born and
Oppenheimer (BO) approximation, and using modern concepts from DFT
\cite{HK:64,KS:65} and perturbation theory \cite{Baroni-RMP:01}.

Within the BO approximation, the lattice-dynamical
properties of a system are determined by the eigenvalues $E$ and
eigenfunctions $\Phi$ of the Schr\"odinger equation: 
\begin{equation}
 \left(-\sum_{I}\frac{\hbar^{2}}{2M_{I}}
   \frac{\partial^{2}}{\partial\mathbf{R}_{I}^{2}}
   +E_{BO}(\{\mathbf{R}\})\right)\Phi(\{\mathbf{R}\})=E
 \Phi(\{\mathbf{R}\}),\label{eq:nucl-schr}
\end{equation}
where $\mathbf{R}_{I}$ is the coordinate of the $I$-th nucleus,
$M_{I}$ its mass, $\{\mathbf{R}\}$ indicates the set of all the
nuclear coordinates, and $E_{BO}$ is the ground-state energy of a
system of interacting electrons moving in the field of fixed nuclei,
whose Hamiltonian---which acts onto the electronic variables and
depends parametrically upon $\{\mathbf{R}\}$---reads:
\begin{equation}
 H_{BO}(\{\mathbf{R}\})=-\frac{\hbar^{2}}{2m}\sum_{i}
 \frac{\partial^{2}}{\partial\mathbf{r}_{i}^{2}}
 +\frac{e^{2}}{2}\sum_{i\ne
   j}\frac{1}{|\mathbf{r}_{i}-\mathbf{r}_{j}|}+
 \sum_{i}V_{\{\mathbf{R}\}}(\mathbf{\mathbf{\mathbf{r}_{i})}}
 +E_{N}(\{\mathbf{R}\}), \label{eq:el-hamilt}
\end{equation}
$-e$ being the electron charge,
$V_{\{\mathbf{R}\}}(\mathbf{r)}=-\sum_{I}
\frac{Z_{I}e^{2}}{|\mathbf{r}-\mathbf{R}_{I}|}$ is the
electron-nucleus interaction, and $ E_{N}(\{\mathbf{R}\})=\frac{e^{2}}{2}\sum_{I\ne
 J}\frac{Z_{I}Z_{J}}{|\mathbf{R}_{I}-\mathbf{R}_{J}|}$ the
inter-nuclear interaction energy.  The equilibrium geometry of the
system is determined by the condition that the forces acting on
individual nuclei vanish: 
\begin{equation} 
\mathbf{F}_{I}\equiv-\frac{\partial E_{BO}(\{\mathbf{R}\})}{\partial\mathbf{R}_{I}}=0,
\end{equation} 
whereas the vibrational frequencies, $\omega$, are determined by the
eigenvalues of the Hessian of the BO energy, scaled by the nuclear
masses: 
\begin{equation} 
 \mathrm{det} \left|\frac{1}{\sqrt{M_{I}M_{J}}}
   \frac{\partial^{2}E_{BO}(\{\mathbf{R}\})}{\partial\mathbf{R}_{I}\partial\mathbf{R}_{J}}
   -\omega^{2}\right|=0.\label{eq:secular-frequency}
\end{equation}

The calculation of the equilibrium geometry and vibrational properties
of a system thus amounts to computing the first and second derivatives
of its BO energy surface. The basic tool to accomplish this goal is
the Hellmann-Feynman (HF) theorem \cite{Hellmann,Feynman}, which
leads to the following expression for the forces: 
\begin{equation}
 \mathbf{F}_{I}=-\int n_{\{\mathbf{R}\}}(\mathbf{r})
 \frac{\partial
   V_{\{\mathbf{R}\}}(\mathbf{r})}{\partial\mathbf{R}_{I}}d\mathbf{r}
 -\frac{\partial
   E_{N}(\mathbf{R})}{\partial\mathbf{R}_{I}},\label{eq:HF-force}
\end{equation}
where $n_{\{\mathbf{R}\}}(\mathbf{r})$ is the ground-state electron charge
density corresponding to the nuclear configuration $\{\mathbf{R}\}$.
The Hessian of the BO energy surface appearing in Eq.\ \eqref{eq:secular-frequency}
is obtained by differentiating the HF forces with respect to nuclear
coordinates: 
\begin{eqnarray}
 \frac{\partial^2E_{BO}(\{\mathbf{R}\})}{\partial\mathbf{R}_I\partial\mathbf{R}_J}
 &\equiv& -\frac{\partial\mathbf{F}_{I}}{\partial\mathbf{R}_{J}} \\ 
 &=& \int\frac{\partial
   n_{\{\mathbf{R}\}}(\mathbf{r})}{\partial\mathbf{R}_{J}}
 \frac{\partial
   V_{\{\mathbf{R}\}}(\mathbf{r})}{\partial\mathbf{R}_{I}}d\mathbf{r}
 \\ &&\quad\quad\quad
 +\int n_{\{\mathbf{R}\}}(\mathbf{r})
 \frac{\partial^{2}V_{\{\mathbf{R}\}}(\mathbf{r})}{\partial 
   \mathbf{R}_I\partial \mathbf{R}_J}d\mathbf{r}+
 \frac{\partial^{2} E_{N}(\{\mathbf{R}\})}{\partial
   \mathbf{R}_I\partial \mathbf{R}_J}.
\label{eq:hessian}
\end{eqnarray}

Eq.\ \eqref{eq:hessian} states that the calculation of the Hessian
of the BO energy surfaces requires the calculation of the ground-state
electron charge density, $n_{\{\mathbf{R}\}}(\mathbf{r})$, as well as
of its \emph{linear response} to a distortion of the nuclear geometry,
${\partial\mathbf{n}_{\{\mathbf{R}\}}(\mathbf{r})/\partial\mathbf{R}_{I}}$.
This fundamental result was first stated in the late sixties by De
Cicco and Johnson \cite{P.D.DeCicco04291969} and by Pick, Cohen,
and Martin \cite{PhysRevB.1.910}. The Hessian matrix is usually called
the matrix of the \textit{inter-atomic force constants} (IFC).
For a crystal, we can write:
\begin{equation}
 C_{ss'}^{\alpha\alpha'}(\mathbf{R}-\mathbf{R}')=
 \frac{\partial^{2}E_{BO}( \{ \mathbf{R} \} )}
 {\partial u_{s}^{\alpha}(\mathbf{R})\partial
   u_{s'}^{\alpha'}(\mathbf{R}')},
 \label{eq:IFC}
\end{equation}
where $u_{s}^{\alpha}(\mathbf{R})$ is the $\alpha$-th Cartesian
components of the displacement of the $s$-th atom of the crystal
unit cell located at lattice site $\mathbf{R}$, and translational
invariance shows manifestly in the dependence of the IFC matrix on
$\mathbf{R}$ and $\mathbf{R}'$ through their difference only.

\subsection{Density-functional perturbation theory}
\label{sec:linear-response}

We have seen that the electron-density linear response of a system
determines the matrix of its IFCs, Eq.\ \eqref{eq:hessian}. Let us see
now how this response can be obtained from DFT. The procedure
described in the following is usually referred to as
density-functional perturbation theory \cite{BGT:87,Baroni-RMP:01}.

In order to simplify the notation and make the argument more general,
we assume that the external potential acting on the electrons is a
differentiable function of a set of parameters, $\boldsymbol{\lambda}\equiv\{\lambda_{i}\}$
($\lambda_{i}\equiv\mathbf{R}_{I}$ in the case of lattice dynamics).
According to the HF theorem, the first and second derivatives of the
ground-state energy read:\begin{eqnarray}
\frac{\partial E}{\partial\lambda_{i}} & = & \int\frac{\partial V^{\boldsymbol{\lambda}}(\mathbf{r})}{\partial\lambda_{i}}n^{\boldsymbol{\lambda}}(\mathbf{r})d\mathbf{r},\label{eq:dfpt-1}\\
\frac{\partial^{2}E}{\partial\lambda_{i}\partial\lambda_{j}} & = & \int\frac{\partial^{2}V^{\boldsymbol{\lambda}}(\mathbf{r})}{\partial\lambda_{i}\partial\lambda_{j}}n^{\boldsymbol{\lambda}}(\mathbf{r})d\mathbf{r}+\int\frac{\partial n^{\boldsymbol{\lambda}}(\mathbf{r})}{\partial\lambda_{i}}\frac{\partial V^{\boldsymbol{\lambda}}(\mathbf{r})}{\partial\lambda_{j}}d\mathbf{r}.\label{eq:dfpt-2}\end{eqnarray}

In DFT the electron charge-density distribution, $n^{\boldsymbol{\lambda}}$,
is given by:\begin{equation}
n^{\boldsymbol{\lambda}}(\mathbf{r})=2\sum_{n=1}^{N/2}|\psi_{n}^{\boldsymbol{\lambda}}(\mathbf{r})|^{2},\label{eq:n_of_r}\end{equation}
where $N$ is the number of electrons in the system (double degeneracy
with respect to spin degrees of freedom is assumed), the single-particle
orbitals, $\psi_{n}^{\boldsymbol{\lambda}}(\mathbf{r})$, satisfy
the Kohn-Sham (KS) Schr\"odinger equation:\begin{equation}
\left(-\frac{\hbar^{2}}{2m}\frac{\partial^{2}}{\partial\mathbf{r}^{2}}+V_{SCF}^{\boldsymbol{\lambda}}(\mathbf{r})\right)\psi_{n}^{\boldsymbol{\lambda}}(\mathbf{r})=\epsilon_{n}^{\boldsymbol{\lambda}}\psi_{n}^{\boldsymbol{\lambda}}(\mathbf{r}),\label{eq:KSeq}\end{equation}
and the \emph{self-consistent} potential, $V_{SCF}^{\boldsymbol{\lambda}}$,
is given by:\begin{equation}
V_{SCF}^{\boldsymbol{\lambda}}=V^{\boldsymbol{\lambda}}+e^{2}\int\frac{n^{\boldsymbol{\lambda}}(\mathbf{r}')}{|\mathbf{r}-\mathbf{r}'|}d\mathbf{r}'+\mu_{XC}[n^{\boldsymbol{\lambda}}](\mathbf{r}),\label{eq:Vscf}\end{equation}
where $\mu_{XC}$ is the so-called \emph{exchange-correlation} (XC)
potential \cite{KS:65}. The electron-density response, ${\partial n_{\boldmath\lambda}(\mathbf{r})/\partial\lambda_{i}}$,
appearing in Eq.\ \eqref{eq:dfpt-2} can be evaluated by linearizing
Eqs.\ \eqref{eq:n_of_r}, \eqref{eq:KSeq}, and \eqref{eq:Vscf}
with respect to wave-function, density, and potential variations,
respectively. Linearization of Eq.\ \eqref{eq:n_of_r} leads to:
\begin{equation}
n'(\mathbf{r})=4\mathrm{Re}\sum_{n=1}^{N/2}\psi_{n}^{*}(\mathbf{r})\psi'_{n}(\mathbf{r}),\label{eq:dn}\end{equation}
where the prime symbol (as in $n'$) indicates differentiation with
respect to one of the $\lambda$'s. The super-script $\boldsymbol{\lambda}$
has been omitted in Eq.\ \eqref{eq:dn}, as well as in any subsequent
formulas where such an omission does not give rise to ambiguities.
Since the external potential (both unperturbed and perturbed) is real,
KS eigenfunctions can be chosen to be real, and the sign of complex
conjugation, as well as the prescription to keep only the real part,
can be dropped in Eq. \eqref{eq:dn}.

The variation of the KS orbitals, $\psi'_{n}(\mathbf{r})$, is obtained
by standard first-order perturbation theory \cite{Messiah}: \begin{equation}
(H_{SCF}^{\circ}-\epsilon_{n}^{\circ})|\psi'_{n}\rangle=-(V'_{SCF}-\epsilon'_{n})|\psi_{n}^{\circ}\rangle,\label{eq:linear}\end{equation}
where $H_{SCF}^{\circ}=-\frac{\hbar^{2}}{2m}\frac{\partial^{2}}{\partial\mathbf{r}^{2}}+V_{SCF}^{\circ}(\mathbf{r})$
is the unperturbed KS Hamiltonian, \label{eq:dVscf}, \begin{equation}
V'_{SCF}(\mathbf{r})=V'(\mathbf{r})+\int\kappa(\mathbf{r},\mathbf{r}')n'(\mathbf{r}')d\mathbf{r}'\label{eq:V-prime}\end{equation}
is the first-order correction to the self-consistent potential, Eq.
\eqref{eq:Vscf}, $\kappa(\mathbf{r},\mathbf{r}')=\frac{e^{2}}{|\mathbf{r}-\mathbf{r}'|}+\frac{\delta\mu_{XC}(\mathbf{r})}{\delta n(\mathbf{r}')}$
is the \emph{Hartree-plus-XC kernel}, and $\epsilon'_{n}=\langle\psi_{n}^{\circ}|V'_{SCF}|\psi_{n}^{\circ}\rangle$
is the first order variation of the KS eigenvalue, $\epsilon_{n}$.
Equations (\ref{eq:dn}--\ref{eq:V-prime}) form a set of self-consistent
equations for the perturbed system completely analogous to the KS
equations in the unperturbed case---Eqs.\  \eqref{eq:n_of_r}, \eqref{eq:KSeq},
and \eqref{eq:Vscf} ---with the KS eigenvalue equation, Eq.\ \eqref{eq:KSeq},
being replaced by a linear system, Eq. \eqref{eq:linear}. The computational
cost of the determination of the density response to a single perturbation
is of the same order as that needed for the calculation of the unperturbed
ground-state density.

The above discussion applies to insulators, where there is a finite
gap. In metals a finite density of states occurs at the Fermi energy,
and a change in the orbital occupation number may occur upon the application
of an infinitesimal perturbation. The modifications of DFPT needed
to treat the linear response of metals are discussed in Refs. \cite{PhysRevB.51.6773,Baroni-RMP:01}.

\subsection{Interatomic force constants and phonon band interpolation}

The above discussion indicates that the primary physical ingredient of
a lattice-dynamical calculation is the IFC matrix,
Eq. \eqref{eq:hessian}, from which vibrational frequencies can be
obtained by solving the secular problem,
Eq. \eqref{eq:secular-frequency}. That phonon frequencies can be
classified according to a well defined value of the crystal momentum
$\mathbf{q}$ follows from the translational invariance of the IFC
matrix. Because of this, the IFC matrix can be Fourier analyzed to
yield the so called \emph{dynamical matrix}, prior to
diagonalization:
\begin{equation}
 \widetilde{C}_{st}^{\alpha\beta}(\mathbf{q})=
 \sum_{\mathbf{R}}C_{st}^{\alpha\beta}(\mathbf{R})
 \mathrm{e}^{-i\mathbf{q}\cdot\mathbf{R}},\label{eq:Cofq}
\end{equation}
and the squared vibrational frequencies, $\omega(\mathbf{q},\nu)^{2}$,
are the eigenvalues of the $3n\times3n$ dynamical
matrix:
\begin{equation}
 D_{st}^{\alpha\beta}(\mathbf{q})=
 \frac{1}{\sqrt{M_{s}M_{t}}}\widetilde{C}_{st}^{\alpha\beta}(\mathbf{q}),
 \label{eq:dynmat_def}
\end{equation}
$n$ being the number of atoms in the unit cell. The direct computation
of the IFCs is unwieldy because it requires the calculation of the
crystal electronic linear response to a localized perturbation (the
displacement of a single atom or atomic plane), which would in turn
break the translational symmetry of the system, thus requiring the use
of computationally expensive large unit cells
\cite{martin-book,Alfe,Parlinski}.  The IFCs are instead more
conveniently calculated in Fourier space, which gives direct access to
the relevant $\mathbf{q}$-dependent dynamical matrices
\cite{Baroni-RMP:01}. Because of translational invariance, the linear
response to a \emph{monochromatic} perturbation, i.e. one with a
definite wave-vector $\mathbf{q}$, is also monochromatic, and all
quantities entering the calculation can be expressed in terms of
lattice-periodic quantities \cite{Baroni-RMP:01}. As a consequence,
vibrational frequencies can be calculated at any wave-vector in the
BZ, without using any supercells, with a numerical effort that is
independent of the phonon wave-length and comparable to that of a
single ground-state calculation for the unperturbed system.

The accurate calculation of sums (integrals) of lattice-dynamical
properties over the BZ (such as those appearing in the QHA formulation
of the thermodynamics of crystals in Sec. \ref{sec:QHA}) requires
sampling the integrand over a fine grid of points. This may be
impractical in many cases, and suitable interpolation techniques are
therefore called for. The most accurate, and physically motivated,
such technique consists in the calculation of real-space IFCs by
inverse analyzing a limited number of dynamical matrices calculated on
a coarse grid. Dynamical matrices at any arbitrary point in the BZ can
then be inexpensively reconstructed by Fourier analysis of the IFC's
thus obtained. According to the \emph{sampling theorem} by Shannon
\cite{shannon}, if the IFCs are strictly short-range, a finite number
of dynamical matrices, sampled on a correspondingly coarse
reciprocal-space grid, is sufficient to calculate them \emph{exactly}
by inverse Fourier analysis. The IFCs thus obtained can then be used
to calculate exactly the dynamical matrices at any wave-vector not
included in the original reciprocal-space grid. In the framework of
lattice-dynamical and band-structure calculations this procedure is
usually referred to as \emph{Fourier interpolation}. Of course, IFCs
are never strictly short-range, and Fourier interpolation is in
general a numerical approximation, subject to so-called
\emph{aliasing} errors, whose magnitude and importance have to be
checked on a case-by-case basis.

Let us specialized to the case of a crystal, in which lattice vectors
${\bf R}$ are generated by primitive vectors ${\bf a}_{1}$, ${\bf a}_{2}$,
${\bf a}_{3}$: $\mathbf{R}_{lmn}=l{\bf a}_{1}+m{\bf a}_{2}+n{\bf a}_{3}$,
with $l,m,n$ integer numbers. The reciprocal lattice vectors ${\bf G}$
are generated in an analogous way by vectors ${\bf b}_{1}$, ${\bf b}_{2}$,
${\bf b}_{3}$, such that 
\begin{equation}
 {\bf a}_{i}\cdot{\bf b}_{j}=2\pi\delta_{ij}. \label{eq:biortho}
\end{equation}
Correspondingly we consider a symmetry-adapted uniform grid of \textbf{q}-vectors:
\begin{equation}
 \mathbf{q}_{pqr}=\frac{p}{N_{1}}{\bf b}_{1}+
 \frac{q}{N_{2}}{\bf b}_{2}+\frac{r}{N_{3}}{\bf b}_{3},
\end{equation}
where $p,q,r$ are also integers. This grid spans the reciprocal
lattice of a supercell of the original lattice, generated by primitive
vectors $N_{1}{\bf a}_{1}$, $N_{2}{\bf a}_{2}$, $N_{3}{\bf
 a}_{3}$. Since wave-vectors differing by a reciprocal-lattice vector
are equivalent, all values of $p,q,r$ differing by a multiple of
$N_{1},N_{2},N_{3}$ respectively, are equivalent. We can then restrict
our grid to $p\in[0,N_{1}-1]$, $q\in[0,N_{2}-1${]}, and
$r\in[0,N_{3}-1]$.  The $\mathbf{q}_{pqr}$ grid thus contains
$N_{1}\times N_{2}\times N_{3}$ uniformly spaced points and spans the
parallelepiped generated by ${\bf b}_{1}$, ${\bf b}_{2}$, ${\bf
 b}_{3}$. It is often convenient to identify wave-vectors with
integer labels spanning the $\left[-\frac{N}{2},\frac{N}{2}-1\right]$
range, rather than $[0,N-1].$ Negative indeces can be folded to
positive values using the periodicity of discrete Fourier transforms.

Once dynamical matrices have been calculated on the ${\bf q}_{hkl}$
grid, IFCs are easily obtained by (discrete) fast-Fourier transform
(FFT) techniques: 
\begin{eqnarray*}
 C_{st}^{\alpha\beta}(\mathbf{R}_{lmn}) & = & 
 \frac{1}{N_{1}N_{2}N_{3}}\sum_{pqr}\widetilde{C}_{st}^{\alpha\beta}(\mathbf{q}_{pqr})
 \mathrm{e}^{i\mathbf{q}_{pqr}\cdot\mathbf{R}_{lmn}} \\
 & = & \frac{1}{N_{1}N_{2}N_{3}}\sum_{pqr}
 \widetilde{C}_{st}^{\alpha\beta}(\mathbf{q}_{pqr})
 \mathrm{e}^{i2\pi\left(\frac{lp}{N_{1}}+\frac{mq}{N_{2}}+\frac{nr}{N_{3}}\right)},
\end{eqnarray*}
where the bi-orthogonality of the real- and reciprocal-space primitive
vectors, Eq. \eqref{eq:biortho}, is used to get $\mathbf{q}_{pqr}
\cdot \mathbf{R}_{lmn}=2\pi \left(\frac{lp}{N_{1}} + \frac{mq}{N_{2}}
 + \frac{nr}{N_{3}} \right)$. The IFCs thus obtained can be used to
calculate dynamical matrices at wave-vectors not originally contained
in the reciprocal-space grid.  This can be done directly wave-vector
by wave-vector, or by FFT techniques, by padding a conveniently large
table of IFCs with zeroes beyond the range of those calculated from
Fourier analyzing the original coarse reciprocal-space grid.

\section{Computer codes}

In order to implement the QHA from first principles, one needs to
compute the complete phonon dispersion of a crystal for different
values of the crystal volume. This can be done within DFT by the \emph{direct}
or \emph{frozen phonon} method, or by the \emph{linear response} method
\cite{Baroni-RMP:01,martin-book}. The former does not require the
use of specialized software beside that needed to perform standard
ground-state DFT calculations, but is computationally more demanding.
Some software tools that help analyze the output of standard DFT code
to produce real-space IFC's and, from these, reciprocal-space dynamical
matrices are available \cite{Alfe,Parlinski}. As for the linear-response
approach, two widely known general-purpose packages exist, \textsc{Quantum
ESPRESSO} \cite{QE} and \emph{ABINIT} \cite{Abinit}. In the following
we briefly describe the former, as well as another code, \emph{QHA},
that can be used as a post-processing tool to perform QHA calculations
starting from lattice-dynamical calculations performed with many different
methods (semi-empirical as well as first-principles, frozen-phonon,
as well as DFPT).

\subsection{Quantum ESPRESSO}

\textsc{Quantum ESPRESSO}\emph{ (opEn Source Package for Research in
 Electronic Structure, Simulation, and Optimization}) is an
integrated suite of computer codes for electronic-structure
calculations and materials modeling, based on DFT, plane waves,
pseudopotentials (norm-conserving and ultrasoft) and all-electron
Projector-Augmented-Wave potentials \cite{QE}. It is freely available
under the terms of the GNU General Public License.\textsc{ Quantum
 ESPRESSO} is organized into packages. For the purposes of
lattice-dynamical calculations and QHA applications, the two most
relevant ones are \texttt{PWscf} and \texttt{PHonon}. The former
produces the self-consistent electronic structure and all related
computations (forces, stresses, structural optimization, molecular
dynamics). The latter solves the DFPT equations and calculates
dynamical matrices for a single wave-vector or for a uniform grid of
wave-vectors; Fourier interpolation can be applied to the results to
produce IFCs up to a pre-determined range in real space. The effects
of macroscopic electric field are separately dealt with using the
known exact results valid in the long-wavelength limit
\cite{Born-Huang}. Both the electronic contribution to the dielectric
tensor, $\epsilon_{\infty}$, and the effective charges $Z^{\star}$ are
calculated by \texttt{PHonon} and taken into account in the calculation
of interatomic force constants.  Once these have been calculated,
phonon modes at any wave-vector can be recalculated in a quick and
economical way. Anharmonic force constants can be explicitly
calculated using the \texttt{D3} code contained in the \texttt{PHonon}
package. The volume dependence of the IFCs needed within the QHA is
simply obtained numerically by performing several phonon (harmonic)
calculations at different volumes of the unit cell.

\subsection{The {\itshape QHA} code}

Once the IFC matrix (or, equivalently, the dynamical matrix over a
uniform grid in reciprocal space) has been calculated, thermodynamical
properties can be easily calculated using the \emph{QHA} code \cite{QHA}.
\emph{QHA} requires in input just a few data: basic information
about the system (such as atomic masses, lattice type) and
a file containing IFCs, stored in an appropriate format. \emph{QHA}
then calculates and several quantities such as the total phonon density
of states (DOS), atom-projected DOS, the isochoric heat capacity,
the Debye temperature, zero-point vibration energy, internal energy,
entropy, mean square displacements for atoms, etc. The DOS is obtained
via the tetrahedron method \cite{tetrahedra}, while integrals over
the frequency are calculated using the Simpson's {}``3/8 rule''.

\section{Applications}

The first investigations of the thermal properties of materials using
\emph{ab initio} phonons and the QHA date back to the early days of
DFPT theory, when the thermal expansivity
of tetrahedrally coordinated semiconductors and insulators was first
addressed \cite{Fleszar,pavone-phd,Pavone_Diamond}. Many other applications
have appeared ever since to metals, hydrides, intermetallic compounds,
surfaces, and to systems and properties of mineralogical and geophysical
interest. Brief reviews of these applications can be found in Refs.
\cite{Baroni-RMP:01,Rickman}; this section contains a more up-to-date
review, with a special attention paid to those applications that are
relevant to the Earth Sciences.

\subsection{Semiconductors and insulators}

One of the most unusual features of tetrahedrally coordinated elemental
and binary semiconductors is that they display a negative thermal
expansion coefficient (TEC) at very low temperature. This finding
prompted the first applications of the QHA to semiconductors, using
first a semi-empirical approach \cite{PhysRevLett.63.290}, and first-principles
techniques in the following \cite{Fleszar,pavone-phd,Pavone_Diamond,Hamdi,Debernardi_96,Gaal,Rignanese,Xie,Grimvall_Si,Marzari,Zimmermann}.
The detailed insight provided by the latter allowed one to trace back
this behavior to the negative Gr\"uneisen parameter in the lowest acoustic
phonon branch and to its flatness that enhances its weight in the
vibrational density of states at low frequency. This behavior is not
observed in diamond at ambient conditions---which in fact does not
display any negative TEC \cite{Pavone_Diamond,Xie}---whereas at pressures
larger than $\sim700\,\mathrm{GPa}$ the softening of the acoustic
Gr\"uneisen parameters determines a negative TEC. The TEC of diamond
calculated in Ref. \cite{Pavone_Diamond} starts deviating from experimental
points at $\mathrm{T\gtrsim600K}$ which was explained in terms of
enhanced anharmonic effects at higher temperature. However, a recent
calculation done with a different XC energy functional (GGA, rather than
LDA) \cite{Marzari} displayed a fairly good agreement with experiments
up to $\mathrm{T=1200K}$, and with results of Monte-Carlo simulations
\cite{Herrero} up to $\mathrm{T=3000K}$. Graphite shows negative
in-plane TEC over a broad temperature range, up to 600K, and the calculated
TEC for graphene is negative up to 2000K \cite{Marzari}. This is
due to a negative Gr\"uneisen parameter of the out-of-plane lattice
vibrations along the $\Gamma M$ and $\Gamma K$ directions (the so
called ZA modes, which plays an important role in the thermal properties
of layered materials, due to the high phonon DOS displayed at low
frequency because of a vanishing sound velocity \cite{ZA-1,ZA-2}).
Such an unusual thermal contraction for carbon fullerenes and nanotubes
was confirmed by molecular dynamics simulations in Ref. \cite{Kwon_PRL}.
The heat capacity of carbon nanotubes was calculated in Ref. \cite{Zimmermann}.
The out-of-plane TEC calculated for graphite \cite{Marzari}
is in poor agreement with experiment. This is not unexpected because
inter-layer binding is mostly due to dispersion forces which
are poorly described by the (semi-) local XC functionals currently
used in DFT calculations.

One of the early achievements of DFT that greatly contributed to its
establishment in the condensed-matter and materials-science
communities was the prediction of the relative stability of different
crystal structures as a function of the applied pressure
\cite{Gaal,Grimvall_Si,Correa,Liu,Isaev_PNAS,Arkady,Arkady_Au}.
Thanks to the QHA, vibrational effects can be easily included in the
evaluation of the crystal free energy, thus allowing for the
exploration of the phase diagram of crystalline solids at finite
temperature.  In Refs. \cite{Gaal,Grimvall_Si}, for instance, the $P-T$
phase diagram for Si and Ge was studied in correspondence to the
$\mathrm{diamond\to\beta\hbox{-Sn}}$ transition. Noticeable changes in
the EOS of ZnSe at finite temperature were shown in \cite{Hamdi}. The
phase boundary between cubic and hexagonal BN has been studied in
Ref. \cite{Hafner} using the QHA with an empirical correction to
account for the leading (explicit) anharmonic effects. Other
applications of the QHA in this area include the low-temperature
portion ot the $P-T$ phase diagram for the diamond $\to$ BC8 phase
transition \cite{Correa} and the sequence of Rhombohedral (223K) $\to$
Orthorhombic (378K) $\to$ Tetragonal (778K) $\to$ Cubic phase
transitions in $\mathrm{BaTiO_{3}}$ \cite{Zhang_BaTiO3} at ambient
pressure.

\subsection{Simple metals}

The QHA has been widely used to investigate
the thermal properties of BCC \cite{Quong,Liu,Debernardi_2001}, FCC
\cite{Debernardi_2001,Tse_Al3Li,Grabowski,Xie_Ni,Narasimhan_Cu,Xie_Ag,Tsuchiya_Au,Wentzcovitch_Pt},
and HCP \cite{Ismail,Renata_Mg} metals. These works generally report
a good agreement with experiments as concerns the calculated lattice
volume, bulk modulus, TEC, Gr\"uneisen parameter, and high-pressure/high-temperature
phase diagram. Some discrepancies in the temperature dependence of
$C_{P}$ and TEC might be connected to the neglect of explicit anharmonic
effects  at high temperatures, as well as due to overestimated cell
volumes when using GGA XC functionals. In Ref. \cite{Grabowski} it
was stressed that implicit quasi-harmonic effects dominate the thermal
properties, being almost two orders of magnitude larger than explicit
anharmonic ones, irrespective of the XC functional adopted.

The QHA has also been an important ingredient in the calculation of
the melting curve of some metals, such as Al \cite{Vocadlo_Al}, Si
\cite{Alfe_Si}, and Ta \cite{Gulseren_Ta,Alfe_Ta}, performed via
thermodynamic integration. The vibrational contribution to the low-temperature
free energy of the crystal phase was shown to be important for lighter
elements (such as Al), whereas it is negligible for heavier ones,
such as Ta. The $P-T$ phase diagram for HCP-BCC Mg has been obtained
in Ref. \cite{Renata_Mg}, where it was shown that a proper account
of lattice vibrations improves the prediction of the transition pressure
at room temperature. Interestingly, in Ref. \cite{Xie_Ag} it was
noticed that in the QHA equation of state (EOS) of Ag there exists
a critical temperature beyond which no volume would correspond to
a vanishing pressure---thus signaling a thermodynamic instability
of the system---and that this temperature is actually rather close
to the experimental melting temperature of Ag. Narasimhan \emph{et
al. }\cite{Narasimhan_Cu} studied the influence of different (LDA
and GGA) functionals on the thermal properties of Cu. The contribution
of lattice vibrations to the phase stability of Li and Sn has been
studied in Refs. \cite{Liu,Pavone_Sn1,Pavone_Sn2}: a proper account
of vibrational effects considerably improves the predictions of the
low-temperature structural properties of a light element such as Li,
which is strongly affected by zero-point vibrations \cite{Liu}. The
large vibrational entropy associated with low-frequency modes stabilizes
the BCC structure of Li \cite{Liu} and $\beta$-Sn \cite{Pavone_Sn1,Pavone_Sn2}
just above room temperature.

\subsection{Hydrides }

One of the best illustrations of the ability of the QHA to account for
the effects of lattice vibrations on the relative stability of
different crystalline phases is provided by iron and palladium
hydrides, FeH and PdH. FeH was synthesized by different experimental
groups \cite{Antonov,Badding,Hirao} and its crystalline structure was
found to be a double hexagonal hexagonal structure (DHCP), contrary to
the results of \emph{ab initio} calculations \cite{Elsasser} that,
neglecting vibrational effects, would rather predict a simple HCP
structure.  The puzzle remained unsolved until free-energy
calculations for FCC, HCP, and DHCP FeH \cite{Isaev_PNAS} showed that
the hydrogen vibrational contribution to the free energy actually
favors the DHCP structure.  This is a consequence of the linear
ordering of H atoms in HCP FeH, which shifts to higher frequencies the
mostly H-like optical band of the system, with respect to the FCC and
DHCP phases. The corresponding increase in the zero-point energy makes
the DHCP structure---which is the next most favored, neglecting
lattice vibrations---the stablest structure at low pressure.The
quantum nature of hydrogen vibrations and its influence on the phase
stability of hydrides was also clearly demonstrated in
\cite{Alavi_JMol,Hu}. First-principles pseudopotential calculations
for PdH have shown that tetrahedrally coordinated H (B3-type PdH) is
energetically favored with respect to octahedrally coordinated H
(B1-type PdH), at variance with experimental findings
\cite{Rowe,Nelin}. The quantum-mechanical behavior of hydrogen
vibrations dramatically affects on the stability of PdH phases at low
temperature, favoring the octahedral coordination of hydrogen atoms in
PdH \cite{Alavi_JMol}. As another example, the QHA does not predict
any $\alpha\to\beta$ (monoclinic to orthorhombic) phase transition in
$\mathrm{Na_{2}BeH_{4}}$ \cite{Hu}, contrary to the conclusions that
were reached from static total-energy calculations.   Overall, the
structural parameters of most alkaline hydrides calculated using the QHA  
turned out to be substantially improved by a proper account of zero-point
vibrations, both  using LDA and GGA XC functionals (more so in the
latter case) \cite{Roma_LiH,Barrera_LiH,Pickett_LiH,Zhang_LiH}.

\subsection{Intermetallics }

The QHA has been also successfully applied to the thermal properties
of intermetallics and alloys. For example, the Gr\"uneisen parameters,
isothermal bulk modulus, TEC, and constant-pressure specific heat
for $\mathrm{Al_{3}Li}$ have been calculated in \cite{Tse_Al3Li}.
The TEC temperature dependence of the technologically important superalloys
B2 NiAl and $\mathrm{L1_{2}}$ Ni$_{3}$Al, as well as $\mathrm{L1_{2}}$
Ir$_{3}$Nb, have been studied in Refs. \cite{Wang_Acta_Mat,Arroyave,Lozovoi,Gornostyrev}.
This is a very significant achievement of QHA, as it makes it possible
very accurate temperature-dependent calculations of the misfit between
lattice parameters of low-temperature FCC/BCC alloy and high-temperature
$\mathrm{L1_{2}}$/B2 phases, which plays a considerable role in the
shape formation of precipitates. It has been found that zero-point
vibrations do not affect the type of structural defects in B2 NiAl,
nor do they change qualitatively the statistics of thermal defects
in B2 NiAl \cite{Lozovoi}. Ozolins\emph{ et al.} \cite{Ozolins}
and Persson \emph{et al.} \cite{Persson} have studied the influence
of vibrational energies on the phase stability in Cu-Au and Re-W alloys,
using a combination of the QHA and of the cluster-variation method.
It turns out that lattice vibrations considerably enhance to the stability
of CuAu intermetallic compounds and Cu-Au alloys with respect to phase
separation \cite{Ozolins}, as well as to the relative stability
of the ordered vs. disordered phases at high temperature \cite{Persson}.

\subsection{Surfaces}

{\it Ab initio} calculations for surfaces coupled with the QHA have been
done for the past 10 years. For example, an anomalous surface thermal
expansion, the so called \emph{surface pre-melting}, has been studied
for a few metallic surfaces, such as $\textrm{Al(001)}$ \cite{Hansen},
$\mathrm{Al(111)}$\cite{Narasimhan_ZPhys}, $\mathrm{Ag(111)}$\cite{Xie_Ag111,Narasimhan_ZPhys,Al-Rawi},
$\mathrm{Rh(001)}$, $\mathrm{Rh(110)}$ \cite{Xie_Rh}, $\mathrm{Mg(10\overline{1}0)}$
\cite{Ismail}, $\mathrm{Be(10\overline{1}0)}$ \cite{Lazzeri} and
$\mathrm{Be(0001)}$ \cite{Pohl_PRL80}. Hansen \emph{et al.} \cite{Hansen}
noticed that the QHA is fairly accurate up to the Debye temperature,
above which \emph{explicit} anharmonic effects, not accounted for
in this approximation, become important. While no peculiar effects
for the surface inter-layer spacing were found for Al(111) \cite{Narasimhan_ZPhys},
for Ag and Rh surfaces it was found that the outermost interlayer
distance, $d_{12}$, is reduced at room temperature, with respect
to its bulk value, whereas it is expanded at high temperatures \cite{Narasimhan_ZPhys,Xie_Ag111,Al-Rawi,Xie_Rh}.
The expansion of $d_{12}$ in the Ag and Rh surfaces, as well as in
Be(0001) \cite{Pohl_PRL80}, is related to the softening of some in-plane
vibrational modes with a corresponding enhancement of their contribution
to the surface free energy. Free energy calculations for $\mathrm{Be(10\overline{1}0)}$
\cite{Lazzeri} and $\mathrm{Mg(10\overline{1}0)}$ \cite{Ismail}
successfully account for the experimentally observed oscillatory behavior
of the interatomic distances. The large contraction of $d_{12}$ in
$\mathrm{Be(10\overline{1}0)}$ was explained in terms of a strong
anharmonicity in the second layer in comparison with the surface layer
(see also \cite{Marzari_Al110}). For $\mathrm{Be(0001)}$ no oscillatory
behavior in inter-layer spacings was observed in \cite{Pohl_PRL80},
but an anomalously large surface thermal expansion does occur.

\subsection{Earth Materials}

The extreme temperature and pressure conditions occurring in the Earth
interior make many geophysically relevant materials properties and
processes difficult, if not impossible, to observe in the laboratory.
Because of this, computer simulation is often a premier, if not unique,
source of information in the Earth sciences. By increasing the pressure,
the melting temperature also increases, so that the temperature range
over which a material behaves as a harmonic solid is correspondingly
expanded, thus making the QHA a very useful tool to investigate materials
properties at Earth-science conditions.

Iron, the fourth most abundant element on Earth and the main
constituent of the Earth core, plays an outstanding role in human life
and civilization.  In Refs. \cite{Kormann,Sha1,Sha2} the
thermodynamics and thermoelastic properties of BCC Fe have been
treated by means of the QHA and finite-temperature DFT. The
temperature dependence of the calculated constant-pressure heat
capacity deviates from experiment at room temperature, but a proper
inclusion of magnetic effects dramatically improves the agreement up
to the Curie temperature \cite{Kormann}. The calculated Debye
temperature and low-temperature isochoric heat capacity $C_{V}$ are in
good agreement with available experimental data. The magnitude and
temperature dependence of the calculated $C_{12}$, $C_{44}$
elastic constants \cite{Sha2} are consistent with experiment
\cite{Elastic_exp1,Elastic_exp2,Elastic_exp3} in the temperature range
from 0K to 1200K at ambient pressure, while $C_{11}$ is
overestimated \cite{Sha2}, likely because of an underestimated
equilibrium volume. The ambient-pressure shear and compressional sound
velocities are consistent with available ultrasonic measurements.  The
$c/a$ ratio of $\varepsilon$-Fe has been studied in \cite{Sha3} up to
temperatures of 6000 K and pressures of 400 GPa by using the QHA,
resulting in good agreement with previous calculations
\cite{Gannarelli} and X-Ray diffraction experiment \cite{Fe_c2a_Exp}.
A combination of experiments and calculations performed within the QHA
was used to show that the FCC and HCP phases of nonmagnetic Fe
\cite{Arkady} can co-exist at very high temperatures and pressures
($\sim6600\,\mathrm{K}$ and $400\,\mathrm{GPa}$), due to quite small
free-energy differences.

B1-type MgO and CaO, $\mathrm{MgSiO_{3}}$ perovskite, the aragonite
and calcite phases of $\mathrm{CaCO_{3}}$, the various polymorphs
of aluminum silicate, $\mathrm{Al_{2}SiO_{5}}$, silica, $\mathrm{SiO_{2}}$
and alumina, $\mathrm{Al_{2}O_{3}}$ are very important constituents
of the Earth's crust and lower mantle. Besides, it is believed that
the Earth's $D''$ layer is mostly composed of post-perovskite $\mathrm{MgSiO_{3}}$,
while $\gamma$-spinel $\mathrm{Mg_{2}SiO_{4}}$ is the dominant mineral
for the lower part of Earth's transition zone. Note that Mg-based
minerals do contain some amount of Fe substituting Mg.\emph{ }The
high-pressure crystalline structure and stability of these phases
are discussed in \cite{Oganov_NATO,Oganov_ZKrist}. Lattice dynamics
and related thermal and elastic properties of B1 MgO have been studied
in Refs. \cite{Strachan,Drummond,Oganov_JChemPhys,Oganov_PRB67,QHA_MgO_Karki,Karki_MgO_Science286,QHA_Wu_JGeolRes113,QHA_Wu_PRB}.
Wentzcovich and \emph{et al. } have introduced a \emph{semi-empirical
ansatz} that allows for an account of explicit anharmonic contributions
to the QHA estimate of various quantities, such as the TEC and $C_{p}$
\cite{QHA_Wu_JGeolRes113,QHA_Wu_PRB}, resulting in a much improved
agreement with experiments. The temperature and pressure dependence
of elastic constants of B1 MgO \cite{QHA_MgO_Karki,Karki_MgO_Science286,Isaak_JGR_95}
calculated within QHA show very good agreement with experimental data
\cite{MgO_Gamma_Exp}. Besides, pressure dependence of \emph{ab initio}
$\:$ compressional and shear sound velocities is in consistent with
seismic observations for the Earth's lower mantle \cite{Karki_MgO_Science286}.
In contrast with these successes, the calculated thermal properties
of the B1 and B2 phases of CaO \cite{Karki_CaO} are inconsistent
with experimental data, and this is most likely due to the too small
lattice parameter predicted by the LDA, as later investigations based
on a GGA XC functional seem to indicate \cite{Zhang_CaO}. 

The thermal properties of $\mathrm{MgSiO_{3}}$ and $\mathrm{Mg_{2}SiO_{4}}$
and the phase transition boundary in these minerals (perovskite$\to$post-perovskite
$\mathrm{MgSiO_{3}}$ and spinel$\to$post-spinel $\mathrm{Mg_{2}SiO_{4}}$)
have been extensively studied\cite{Wentzcovitch_MgSiO3,Oganov_MgSiO3_Nature,Oganov_JChemPhys_122,QHA_Wu_PRB,Yu_GRL,QHA_Wu_JGeolRes113,Yu_EPSL,Ono_EPSL_2005}
due to their great importance for the Earth's $D''$ layer and
lower mantle, respectively. Improved EOS of B1 MgO \cite{QHA_Wu_JGeolRes113},
obtained by means of \emph{renormalized }phonons and QHA, has been
used as a new pressure calibration to re-evaluate the high pressure -- high temperature
phase boundary in MgSiO3 and Mg2SiO4 minerals using experimental data
from \cite{Fei_JGeophysRes_2004,Hirose,Speziale}. 

Alumina, $\mathrm{Al_{2}O_{3}}$, plays an important role in high-pressure
experiments: for example, it serves as a window material for shock-wave
experiments. Cr-doped alumina, ruby, is used as a pressure calibration
material in diamond-anvil-cell experiments. Besides, it is a component
of solid solutions with $\mathrm{MgSiO_{3}}$ polymorphs
that have significantly different  thermal properties from
pure $\mathrm{MgSiO_{3}}$ minerals. Corundum ($\alpha$-$\mathrm{Al_{2}O_{3}}$)
is the most stable phase of alumina at ambient conditions, preceded
by the $\theta$ phase at lower temperature. The energy difference
between the $\theta$ and $\alpha$ phases of alumina is rather small,
and this raised a question as to whether $\alpha$-$\mathrm{Al_{2}O_{3}}$
is stabilized by phonons. Zero-point vibrations stabilize the corundum
phase at low temperatures \cite{Parlinski_Al2O3}, whereas free-energy
calculations show that the $\alpha$ phase can not be stabilized by
phonons only at room temperature. QHA calculations revealed that at
high pressures alumina transforms to $\mathrm{CaIrO_{3}}$- \cite{Oganov_Al2O3_PNAS}
and $\mathrm{U_{2}S_{3}}$-type \cite{PNAS_Umemoto} polymorphs. 

The $P-T$ phase diagram for $\mathrm{Al_{2}SiO_{5}}$ polymorphs (andalusite,
sillimanite, and kyanite) \cite{Winkler} and the thermal properties
of $\mathrm{CaCO_{3}}$ polymorphs (calcite and aragonite) \cite{Catti_CaCO3_1,Catti_CaCO3_2}
have been studied within the QHA using model inter-atomic potentials.
The effect of zero-point vibrations on the equilibrium volume in the
calcite phase was found to be quite important and actually larger
than the thermal expansion at relatively high temperature \cite{Catti_CaCO3_1}.
These calculations \cite{Catti_CaCO3_2} were not able to account
for the experimentally observed \cite{Rao} negative in-plane TEC
in calcite. The heat capacity and entropy calculated for the aragonite
phase substantially deviate from experiment. All these problems can
be possibly traced back to the poor transferability of model inter-atomic
potentials. 

The thermal properties of the $\alpha$-quartz and stishovite phases
of $\mathrm{SiO_{2}}$ have been studied in \cite{Lee_SiO2}. The
heat capacities of both phases were found to be in good agreement
with experimental data \cite{SiO2_exp1,SiO2_exp2}, with the stishovite
phase having a lower capacity below 480K. Interestingly, zero-point
vibration energy of the stishovite phase affects on thermodynamical
properties stronger than in the $\alpha$-quartz phase \cite{Lee_SiO2}.
The $P-T$ phase diagram of $\mathrm{SiO_{2}}$ has been examined in
Refs. \cite{Oganov_SiO2_PRB,Oganov_JChemPhys_122}, with emphasis
on the stishovite$\to$$\mathrm{CaCl_{2}}$$\rightarrow$$\alpha$-$\mathrm{PbO_{2}}$$\to$pyrite
structural changes, resulting in a sequence of transitions that do
not correspond to any observed seismic discontinuities within the
Earth. Further investigations at ultrahigh temperature and pressure
show that $\mathrm{SiO_{2}}$ exhibits a pyrite$\to$cotunnite phase
transition at conditions that are appropriate for the core of gas giants
and terrestrial exoplanets \cite{Umemoto_Science311}.

\section{Conclusions}

The QHA is a powerful conceptual and practical
tool that complements molecular dynamics in the prediction of the
thermal properties of materials not too close to the melting line.
In the specific case of the Earth Sciences, the QHA can provide information
on the behavior of geophysically relevant materials at those geophysically
relevant pressure and temperature conditions that are not (easily)
achieved in the laboratory. Large-scale calculations using the QHA
for geophysical research will require the deployment of a large number
of repeated structure and lattice-dynamical calculations, as well
as the analysis of the massive data generated therefrom. We believe
that this will require the use of dedicated infrastructures that combine
some of the features of massively parallel machines with those of
a distributed network of computing nodes, in the spirit of the grid
computing paradigm. The \textsc{Quantum ESPRESSO} distribution of
computer codes is geared for exploitation on massively parallel machines
up to several thousands of closely coupled processors and is being
equipped with specific tools to distribute lattice-dynamical calculation
over the grid \cite{phgrid}.

\section{Acknowledgments }

The authors wish to thank Renata M. Wentzcovitch for inspiring some
of their research in this field, as well as for a critical reading
of the manuscript. E.I. thanks the Swedish Research Council VR, the
Swedish Foundation for Strategic Research SSF, the MS2E Strategic
Research Center and the G\"oran Gustafsson Foundation for Research in
Natural Sciences and Medicine, as well as the Russian Foundation for
Basic Researches (grant \#07-02-01266) for financial support. 

\bibliographystyle{jphysicsB}
\bibliography{msa}

\end{document}